\documentstyle[12pt]{article}

\begin{document}

\title{Wavelet analysis as a $p$--adic spectral analysis}
\author{S.V.Kozyrev}
\maketitle

\centerline{\it Institute of Chemical Physics, Russian Academy of
Science}

\begin{abstract}
New orthonormal basis of eigenfunctions for the Vladimirov
operator of $p$--adic fractional derivation is constructed. The
map of $p$--adic numbers onto real numbers
($p$--adic change of variable) is considered. $p$--Adic change of
variable maps the Haar measure on $p$--adic numbers onto
the Lebesgue measure on the positive semiline.
$p$--Adic change of variable (for $p=2$)
provides an equivalence between the constructed
basis of eigenfunctions of the Vladimirov operator and the wavelet
basis in $L^2({\bf R}_+)$ generated from the Haar wavelet. This
means that the wavelet analysis can be considered as a $p$--adic
spectral analysis.
\end{abstract}

\section{Introduction}

In the present paper we construct a new orthonormal basis of
eigenfunctions of the Vladimirov operator of $p$--adic fractional
derivation. The example of such a basis one can find in
\cite{VVZ}. Different basises of eigenvectors of the Vladimirov
operator were built in \cite{AlgAnal}, \cite{Kochubei},
\cite{Trudy}. The basis constructed in the present paper consists
of locally constant functions with support on $p$--adic discs.

We also check that the constructed basis of eigenfunctions of the
Vladimirov operator (for $p=2$) is equivalent to the wavelet basis
in $L^2({\bf R}_+)$ generated from the Haar wavelet. This
equivalence is given by $p$--adic change of variables:
the map of $p$--adic numbers onto real
numbers that conserves the measure. This means that the wavelet
analysis can be considered as a $p$--adic harmonic analysis
(decomposition of functions over the eigenfunctions of the
Vladimirov operator of $p$--adic fractional derivation).

For introduction to $p$-adic analysis see \cite{VVZ}. $p$-Adic
analysis and $p$-adic mathematical physics attract great
interest, see \cite{VVZ}, \cite{UMN}, \cite{Hren}. For instance,
$p$-adic models in string theory were introduced, see
\cite{Vstring}, \cite{Freund}, and $p$-adic quantum mechanics
\cite{VV} and $p$--adic quantum gravity \cite{ADFV} were
investigated. $p$--Adic analysis was applied to investigate the
spontaneous breaking of the replica symmetry, cf. \cite{cond-mat},
\cite{PaSu}, \cite{Carlucci}, \cite{Carlucci1}.

Let us make here a brief review of $p$-adic analysis. The field
$Q_p$ of $p$-adic numbers is the completion of the field of
rational numbers  $Q$ with respect to the $p$-adic norm on $Q$.
This norm is defined in the following way. An arbitrary rational
number $x$ can be written in the form $x=p^{\gamma}\frac{m}{n}$
with $m$ and $n$ not divisible by $p$. The $p$-adic norm of the
rational number $x=p^{\gamma}\frac{m}{n}$ is equal to
$|x|_p=p^{-\gamma}$.

The most interesting property of the field of   $p$-adic numbers
is ultrametricity. This means that $Q_p$ obeys the strong triangle
inequality
$$
|x+y|_p \le \max (|x|_p,|y|_p).
$$
We will consider disks in   $Q_p$ of the form $\{x\in Q_p:
|x-x_0|_p\le p^{-k}\}$. For example, the ring $Z_p$ of integer
$p$-adic numbers is the disk $\{x\in Q_p: |x|_p\le 1\}$ which is
the completion of integers with the $p$-adic norm. The main
properties of disks in arbitrary ultrametric space are the
following:

{\bf 1.}\qquad Every point of a disk is the center of this disk.

{\bf 2.}\qquad Two disks either do not intersect or one of these
disks contains  the other.

The $p$-adic Fourier transform $F$ of the function $f(x)$ is
defined as follows
$$
F[f](\xi)=\widetilde{f}(\xi)=\int_{Q_p}\chi(\xi x)f(x)d\mu(x)
$$
Here $d\mu(x)$ is the Haar measure. The inverse Fourier transform
have the form
$$
F^{-1}[\widetilde{g}](x)=\int_{Q_p}\chi(-\xi
x)\widetilde{g}(\xi)d\mu(\xi)
$$
Here $\chi(\xi x)=\exp(2\pi i\xi x)$ is the character of the field
of $p$-adic numbers.  For example, the Fourier transform of the
indicator function $\Omega(x)$ of the disk of radius 1 with center
in zero (this is a function that equals to 1 on the disk and to 0
outside the disk) is the function of the same type:
$$
\widetilde{\Omega}(\xi)=\Omega(\xi)
$$

In the present paper we use the following Vladimirov operator
$D^{\alpha}_x$ of the fractional $p$-adic differentiation, that is
defined  \cite{VVZ}  as
\begin{equation}\label{diff}
D^{\alpha}_x f(x)=F^{-1}\circ|\xi|_p^{\alpha}\circ F [f](x)=
\frac{p^{\alpha}-1}{1-p^{-1-\alpha}}
\int_{Q_p}\frac{f(x)-f(y)}{|x-y|_p^{1+\alpha}}d\mu(y)
\end{equation}
Here $F$ is the ($p$-adic) Fourier transform, the second equality
holds for $\alpha>0$.

For further reading on the subject of  $p$-adic analysis see
\cite{VVZ}.

\section{$p$--Adic spectral analysis}

Let us construct an orthonormal basis of eigenvectors of
the Vladimirov operator.
The following lemma gives the basic technical result.

\bigskip

\noindent {\bf Lemma 1}.\quad {\sl The function
\begin{equation}\label{psi(x)}
\psi(x)=\chi(p^{-1}x)\Omega\left(|x|_p\right)
\end{equation}
is an eigenvector of the Vladimirov operator:
$$
D^{\alpha}\psi(x)=p^{\alpha}\psi(x)
$$
}

\bigskip

\noindent {\bf Remark}.\quad The eigenvalue for $\psi(x)$ is the
same as for the character $\chi(p^{-1}x)$:
$$
D^{\alpha}\chi(p^{-1}x)=p^{\alpha}\chi(p^{-1}x)
$$

\bigskip

\noindent {\it Proof}\qquad
Let us check that $\psi(x)$ is an
eigenvector of the Vladimirov operator. We get
\begin{equation}\label{integral_y}
D^{\alpha}\psi(x)={p^{\alpha}-1 \over  1-p^{-1-\alpha}}
\int {\psi(x)-\psi(y)\over |x-y|_p^{1+\alpha}}d\mu(y)=
$$
$$
={p^{\alpha}-1\over 1-p^{-1-\alpha}}
\int {\chi(p^{-1}x)\Omega\left(|x|_p\right)-
\chi(p^{-1}y)\Omega\left(|y|_p\right)\over |x-y|_p^{1+\alpha}}d\mu(y)=
$$
$$
=\chi(p^{-1}x) {p^{\alpha}-1\over 1-p^{-1-\alpha}}
\int {\Omega\left(|x|_p\right)-
\chi(p^{-1}(y-x))\Omega\left(|y|_p\right)\over |x-y|_p^{1+\alpha}}d\mu(y)
\end{equation}
Let us calculate the integral over $y$ in (\ref{integral_y}) in
two cases.

1) Let $|x|_p\le 1$. In this case the integral in
(\ref{integral_y}) is given by
$$
\int {1-
\chi(p^{-1}(y-x))\Omega\left(|y|_p\right)\over |x-y|_p^{1+\alpha}}d\mu(y)
$$
Using that every point of $p$--adic disk is its center we get
that for $|x|_p\le 1$ we have $\Omega\left(|y|_p\right)=
\Omega\left(|x-y|_p\right)$ and therefore the integral is equal to
$$
\int {1-
\chi(p^{-1}(y-x))\Omega\left(|y-x|_p\right)\over |x-y|_p^{1+\alpha}}d\mu(y)=
\int {1-
\chi(p^{-1}z)\Omega\left(|z|_p\right)\over |z|_p^{1+\alpha}}d\mu(z)
$$

2) Let $|x|_p > 1$. For the integral in (\ref{integral_y}) we get
$$
- {1\over |x|_p^{1+\alpha}}\int_{|y|_p\le 1}
\chi(p^{-1}(y-x))d\mu(y)=0
$$

This proves that $\psi(x)$ is an eigenvector of the Vladimirov operator
with the following eigenvalue
$$
D^{\alpha}\psi(x)=\psi(x) {p^{\alpha}-1\over 1-p^{-1-\alpha}}
\int {1- \chi(p^{-1}z)\Omega\left(|z|_p\right)\over
|z|_p^{1+\alpha}}d\mu(z)=
$$
$$
=\psi(x) {p^{\alpha}-1\over 1-p^{-1-\alpha}}
\left(p^{-1}\sum_{i=0}^{p-1}(1-\chi(p^{-1}i))+
(1-p^{-1})\sum_{\gamma=1}^{\infty}p^{\gamma}p^{-(1+\alpha)\gamma}\right)=
p^{\alpha}\psi(x)
$$
that finishes the proof of the lemma.

\bigskip

The lemma implies
\begin{equation}\label{rescaling}
D^{\alpha}\psi(ax+b)=|a|^{\alpha}_p p^{\alpha}\psi(ax+b)
\end{equation}
One can check that the set of (integrable) functions $\psi(ax+b)$
(for different $a$, $b$) is a complete system in a Hilbert space
$L^2(Q_p)$. Moreover,

\bigskip

\noindent {\bf Theorem 2}.\quad {\sl The set of functions
$\{\psi_{\gamma j n}\}$:
\begin{equation}\label{basis}
\psi_{\gamma j n}(x)=p^{-{\gamma\over 2}} \chi(p^{\gamma-1}j x)
\Omega(|p^{\gamma} x-n|_p),\quad \gamma\in {\bf Z},\quad n\in
Q_p/Z_p,\quad j=1,\dots,p-1
\end{equation}
is an orthonormal basis in $L^2(Q_p)$ of eigenvectors of the
operator $D^{\alpha}$:
\begin{equation}\label{eigenvalues}
D^{\alpha}\psi_{\gamma j n}= p^{\alpha(1-\gamma)}\psi_{\gamma j n}
\end{equation}
The group $Q_p/Z_p$ in (\ref{basis}) is parameterized by
$$
n=\sum_{i=1}^{m}n_i p^{-i},\quad n_i=0,\dots, p-1
$$
}

\bigskip

\noindent {\it Proof}\qquad Consider the scalar product
\begin{equation}\label{pairing}
\langle \psi_{\gamma j n},\psi_{\gamma' j' n'}\rangle=
$$
$$
=\int_{Q_p} p^{-{\gamma\over 2}} \chi(-p^{\gamma-1}jx)
\Omega(|p^{\gamma}x-n|_p) p^{-{\gamma'\over 2}}
\chi(p^{\gamma'-1}j'x) \Omega(|p^{\gamma'}x-n'|_p)d\mu(x)
\end{equation}
Consider $\gamma\le\gamma'$. We have that the product of
indicators is equal to the indicator or zero:
$$
\Omega(|p^{\gamma}x-n|_p)\Omega(|p^{\gamma'}x-n'|_p)=
\Omega(|p^{\gamma}x-n|_p)\Omega(|p^{\gamma'-\gamma}n-n'|_p)
$$
Since $n'\in Q_p/Z_p$ the function
$\Omega(|p^{\gamma'-\gamma}n-n'|_p)$ is non--zero (and equal to
one) for
$$
n'=p^{\gamma'-\gamma}n
$$
We get for (\ref{pairing})
\begin{equation}\label{integral}
\int_{Q_p} p^{-{\gamma\over 2}} \chi(-p^{\gamma-1}jx)
\Omega(|p^{\gamma}x-n|_p) p^{-{\gamma'\over 2}}
\chi(p^{\gamma'-1}j'x)
\Omega(|p^{\gamma}x-n|_p)\Omega(|p^{\gamma'-\gamma}n-n'|_p)d\mu(x)
\end{equation}

Consider $\gamma<\gamma'$. Then for the integral (\ref{integral})
we get
$$
\int_{Q_p} p^{-{\gamma\over 2}}p^{-{\gamma'\over 2}}
\chi(-p^{\gamma-1}jx)
\chi(p^{\gamma'-\gamma-1}j'n)\Omega(|p^{\gamma}x-n|_p)
\Omega(|p^{\gamma'-\gamma}n-n'|_p)d\mu(x)=0
$$
Therefore the scalar product (\ref{pairing}) can be non--zero only
for $\gamma=\gamma'$. For $\gamma=\gamma'$ the integral
(\ref{integral}) is equal to
$$
\int_{Q_p} p^{-\gamma} \chi(-p^{\gamma-1}jx) \chi(p^{\gamma-1}j'x)
\Omega(|p^{\gamma}x-n|_p)\Omega(|n-n'|_p)d\mu(x)
$$
Since $n$, $n'\in Q_p/Z_p$
$$
\Omega(|n-n'|_p)=\delta_{nn'}
$$
we get for (\ref{pairing})
$$
\langle \psi_{\gamma j n},\psi_{\gamma' j'
n'}\rangle=\delta_{\gamma\gamma'}\delta_{nn'}\int_{Q_p}
p^{-\gamma} \chi(p^{\gamma-1}(j'-j)x)
\Omega(|p^{\gamma}x-n|_p)d\mu(x) =$$
$$
=\delta_{\gamma\gamma'}\delta_{nn'}\delta_{jj'}
$$
that proves that the vectors $\psi_{\gamma j n}$ are orthonormal.

To prove that the set of vectors $\{\psi_{\gamma j n}\}$ is an
orthonormal basis (is total in $L^2(Q_p)$) we use the Parsevale
identity.

Since the set of indicators (characteristic functions) of
$p$--adic discs is total in $L^2(Q_p)$ and the set of vectors
$\{\psi_{\gamma j n}\}$ is translationally invariant and invariant
under dilations $x\mapsto p^n x$, $x\in Q_p$, to prove that
$\{\psi_{\gamma j n}\}$ is a complete system it is enough to check
the Parsevale identity for the indicator $\Omega(|x|_p)$.

We have
\begin{equation}\label{completeness}
\langle\Omega(|x|_p),\psi_{\gamma j n}\rangle= p^{-{\gamma\over
2}}\theta(\gamma)\delta_{n0},\qquad \theta(\gamma)=0,\gamma\le
0,\quad \theta(\gamma)=1,\gamma\ge 1
\end{equation}
Formula (\ref{completeness}) implies the Parsevale identity for
$\Omega(|x|_p)$:
$$
\sum_{\gamma j n}|\langle\Omega(|x|_p),\psi_{\gamma j
n}\rangle|^2=\sum_{\gamma=1}^{\infty}(p-1)p^{-\gamma}=1=
|\langle\Omega(|x|_p),\Omega(|x|_p)\rangle|^2
$$
that proves that $\{\psi_{\gamma j n}\}$ is an orthonormal basis
in $L^2(Q_p)$.

Formula (\ref{rescaling}) implies that the basis $\{\psi_{\gamma j
n}\}$ is an orhtonormal basis of eigenvectors of the operator
$D^{\alpha}$ with eigenvalues (\ref{eigenvalues}). This finishes
the proof of the theorem.

\section{Wavelet interpretation}

Let us discuss the connection between the constructed basis
$\{\psi_{\gamma j n}\}$ and the basis of wavelets in the space
of quadratically integrable functions $L^2({\bf R}_+)$ on positive semiline.
The wavelet basis in $L^2({\bf R}_+)$ is a basis given by shifts and
dilations of the so called mother wavelet function, cf.
\cite{wavelets}. The simplest example of such a function is the
Haar wavelet
\begin{equation}\label{haar}
\Psi(x)=\chi_{[0,\frac{1}{2}]}(x)- \chi_{[\frac{1}{2},1]}(x)
\end{equation}
(difference of two characteristic functions).

The wavelet basis in $L^2({\bf R})$ (or basis of multiresolution
wavelets) is the basis
\begin{equation}\label{basis1}
\Psi_{\gamma  n}(x)=2^{-{\gamma\over 2}}\Psi(2^{-\gamma}x-n),\quad
\gamma\in {\bf Z},\quad n\in {\bf Z}
\end{equation}

Consider the  $p$--adic change of variables, i.e. the onto map
$$
\rho:Q_p \to {\bf R}_+
$$
\begin{equation}\label{change}
\rho:\sum_{i=\gamma}^{\infty} a_i p^{i} \mapsto
\sum_{i=\gamma}^{\infty} a_i p^{-i-1},\quad a_i=0,\dots,p-1,\quad
\gamma \in {\bf Z}
\end{equation}
This map is not a one--to--one map. The map $\rho$ is continuous
and moreover

\bigskip

\noindent{\bf Lemma 3}.\qquad {\sl
The map $\rho$ satisfies the Holder inequality
\begin{equation}\label{norm}
|\rho(x)-\rho(y)| \le |x-y|_p
\end{equation}
}

\medskip

\noindent{\it Proof}\qquad
Consider
$$
x=\sum_{i=\alpha}^{\infty} x_i p^{i},\qquad
y=\sum_{i=\beta}^{\infty} y_i p^{i},
$$
where $\alpha\le\beta$. Then
$$
\rho(x)-\rho(y)=\sum_{i=\alpha}^{\beta-1} x_i p^{i}+
\sum_{i=\beta}^{\infty} (x_i-y_i) p^{i}
$$
Consider the following two cases:

1) Let $\alpha<\beta$. Then
$$
|\rho(x)-\rho(y)|\le (p-1)\sum_{i=\alpha}^{\infty} p^{-i-1}=|x-y|_p
$$

2) Let $\alpha=\beta$. Then $|x-y|_p=p^{-\gamma}$, $\gamma>\alpha$.
$$
\rho(x)-\rho(y)=\sum_{i=\gamma}^{\infty}(x_i-y_i) p^{-i-1}
$$
$$
|\rho(x)-\rho(y)|\le (p-1)\sum_{i=\gamma}^{\infty} p^{-i-1}=
p^{-\gamma}=|x-y|_p
$$
that finishes the proof of the lemma.

\bigskip

The following map is a one--to--one map:
$$
\rho:Q_p/Z_p \to {\bf N}
$$
where ${\bf N}$ is a set of natural numbers including zero.

Here $Q_p/Z_p$ is a group (with respect to addition modulo 1)
of numbers of the form
$$
x=\sum_{i=\gamma}^{-1} x_i p^{i}
$$

\bigskip

\noindent{\bf Lemma 4}.\qquad {\sl
For $n\in Q_p/Z_p$ and $m,k\in {\bf Z}$ the map $\rho$
satisfies the conditions
\begin{equation}\label{in}
\rho: p^m n+p^k Z_p\to p^{-m} \rho(n)+[0,p^{-k}]
\end{equation}
\begin{equation}\label{out}
\rho:Q_p\backslash \{p^m n+p^k Z_p\}\to
{\bf R}_+\backslash\{ p^{-m} \rho(n)+[0,p^{-k}]\}
\end{equation}
up to a finite number of points.}

\medskip

\noindent{\it Proof}\qquad We consider for simplicity the case
$\rho_0=\rho$ and $k=0$. Consider $n\in Q_p/Z_p$:
$$
n=\sum_{i=\gamma}^{-1} n_i p^{i}
$$
$$
n-1=\sum_{i=\gamma}^{-1} n_i p^{i}+\sum_{i=0}^{\infty} (p-1) p^{i}
$$
Then
$$
\rho(n)=\sum_{i=\gamma}^{-1} n_i p^{-i-1}
$$
$$
\rho(n-1)= \rho\left(\sum_{i=\gamma}^{-1} n_i p^{i}+
\sum_{i=0}^{\infty} (p-1) p^{i}\right)= \sum_{i=\gamma}^{-1} n_i
p^{-i-1}+\sum_{i=0}^{\infty} (p-1) p^{-i-1}=\rho(n)+1
$$
Application of (\ref{norm}) for $y=n$, $y=n-1$ proves that
$n+Z_p$ maps into $\rho(n)+[0,1]$.
Since the map $Q_p/Z_p\to {\bf N}$ is one--to--one this proves the lemma.

\bigskip

\noindent{\bf Lemma 5}.\qquad {\sl
The map $\rho$ maps the Haar measure $\mu$ on $Q_p$ onto the Lebesgue
measure $l$ on ${\bf R}_+$: for measurable subspace $X\subset Q_p$
$$
\mu(X)=l(\rho(X))
$$
or in symbolic notations
$$
\rho:d\mu(x)\mapsto dx
$$
}

\medskip

\noindent{\it Proof}\qquad
Lemma 4 implies that balls in $Q_p$ map onto closed intervals  отрезки in
${\bf R}_+$ with conservation of measure.
The map $\rho:Q_p\to {\bf R}_+$ is surjective and moreover
nonintersecting balls map onto intervals thaat do not intersect
or have intersection of the measure zero (by lemma 4).
This proves the lemma.

\bigskip

Therefore the corresponding map
$$
\rho^*:L^2({\bf R}_+)\to L^2({Q_p})
$$
\begin{equation}\label{rhostar}
\rho^*f(x)=f(\rho(x))
\end{equation}
is an unitary operator.

Lemma 4 implies the following:

\bigskip

\noindent{\bf Lemma 6}.\qquad {\sl
The map $\rho$ maps the Haar wavelet
(\ref{haar}) onto the function (\ref{psi(x)}) (for $p=2$):
\begin{equation}\label{mapformother}
\rho^*:\Psi(x)\mapsto \psi(x)
\end{equation}
(\ref{mapformother}) is an equality in  $L^2$:
on the set of zero measure  (\ref{mapformother}) may not be true.
}

\bigskip

Moreover, we have the following theorem:

\bigskip

\noindent {\bf Theorem 7}.\quad {\sl For $p=2$ the map $\rho$
maps the orthonormal basis of wavelets in $L^2({\bf R}_+)$
(\ref{basis1}) (generated from the Haar wavelet) onto the basis
(\ref{basis}) of eigenvectors of the Vladimirov operator:
\begin{equation}\label{mapofbasis}
\rho^*:\Psi_{\gamma \rho(n)}(x)\mapsto (-1)^{n} \psi_{\gamma
1 n}(x)
\end{equation}
}

\medskip

\noindent {\it Proof}\qquad
We have
\begin{equation}\label{dilation}
2^{-\gamma}\rho(x)=\rho(2^{\gamma}x)
\end{equation}
Lemma 4 implies for $\rho(n)\in {\bf N}$
\begin{equation}\label{shift}
\chi_{[0,1]}(\rho(2^{\gamma}x)-\rho(n))=\chi_{[0,1]}(\rho(2^{\gamma}x-n))
\end{equation}
(\ref{shift}) is true almost everywhere. Analogously
\begin{equation}\label{shift1}
\chi_{[0,{1\over 2}]}(\rho(2^{\gamma}x)-\rho(n))=
\chi_{[0,{1\over 2}]}(\rho(2^{\gamma}x-n))
\end{equation}

Formulas (\ref{basis1}), (\ref{dilation}),
(\ref{shift}), (\ref{shift1}) imply
$$
\Psi_{\gamma  \rho(n)}(\rho(x))= 2^{-{\gamma\over
2}}\Psi(2^{-\gamma}\rho(x)-\rho(n))=
$$
$$
=2^{-{\gamma\over 2}}\Psi(\rho(2^{\gamma}x-n))= 2^{-{\gamma\over
2}}\psi(2^{\gamma}x-n)
$$
The last equality follows from (\ref{rhostar}).

Formula (\ref{basis}) implies that for $p=2$
$$
\psi_{\gamma 1 n}(x)=2^{-{\gamma\over 2}} \chi(2^{\gamma-1} x)
\Omega(|2^{\gamma} x-n|_2)=(-1)^n 2^{-{\gamma\over
2}}\psi(2^{\gamma}x-n)
$$
that proves (\ref{mapofbasis}) and finishes the proof of the
theorem.

\bigskip

We get that after the $p$--adic change of variables (\ref{change})
the wavelet analysis becomes the $p$--adic spectral analysis
(expansion of a function over the eigenfunctions of the
Vladimirov operator of $p$--adic derivation).

Using this interpretation we will call the basis (\ref{basis}) the
wavelet basis (or $p$--adic wavelet basis).

Using map (\ref{rhostar}) it is possible to define the action of
the Vladimirov operator in $L^2({\bf R}_+)$ by the formula
$$
\partial_p^{\alpha}f(x)=\rho^{*-1}D^{\alpha}\rho^{*}f(x)
$$
(let us note that $D^{\alpha}$ and $\rho^{*}$ depend on $p$).

Ne can see that
\begin{equation}\label{real}
\partial_p^{\alpha} f(x)=
\frac{p^{\alpha}-1}{1-p^{-1-\alpha}}
\int_{0}^{\infty}\frac{f(x)-f(y)}{|\rho^{-1}(x)-\rho^{-1}(y)|_p^{1+\alpha}}dy
\end{equation}
where $\rho^{-1}$ is the inverse map to $\rho$.
Since $\rho$ is not one--to--one map the map $\rho^{-1}$
is ambiguous but ambiguity is concentrated on the set of zero measure
that makes definition (\ref{real}) correct.

\vspace{5mm} {\bf Acknowledgements}

\vspace{3mm} The author is grateful to V.S.Vladimirov,
I.V.Volovich, V.A.Avetisov, A.H.Bikulov, V.P.Mikhailov and A.K.Guschin
for discussion. This
work was partially supported by INTAS 9900545 and RFFI 990100866
grants.


\begin{thebibliography}{99}
\bibitem{VVZ}\  V.S.Vladimirov,  I.V.Volovich,
Ye.I.Zelenov,  $p$-Adic analysis and mathematical physics. World
Scientific, Singapore, 1994. Russian Edition: Moscow, Nauka, 1994
\bibitem{AlgAnal} \ V.S.Vladimirov, On the spectra of of certain
pseudodifferential operators over the field of $p$--adic numbers,
Algebra i Analiz, vol.2(1990), pp.107--124
\bibitem{Kochubei}\ A.N.Kochubei, Additive and multiplicative
fractional differentiations over the field of $p$--adic numbers,
in ''$p$--Adic Functional Analysis'', Lect. Notes Pure Appl.
Math., vol. 192, New York: Dekker, 1997, pp.275--280
\bibitem{Trudy} \ V.S.Vladimirov, Ramified characters of Idele
groups of one--class quadratic fields, Proc. of the Steklov
Institute of Mathematics, vol.224(1999), pp.107--114
\bibitem{UMN} \ V.S.Vladimirov, Generalized functions over the
field of $p$--adic numbers, Usp. Mat. Nauk, vol.43(1989), No.5
pp.17--53
\bibitem{Hren} \ A.Khrennikov, $p$--Adic valued distributions
in mathematical physics, Kluwer Academic Publ., Dordrecht, 1994
\bibitem{Vstring} \ I.V.Volovich, $p$--Adic String, Class. Quantum Gravity,
4(1987)L83-L87
\bibitem{Freund} \  Freund P.G.O., Olson M., Nonarchimedean strings,
Phys. Lett. B, 1987, Vol.199, p.186
\bibitem{VV} \  Vladimirov V.S.,  Volovich I.V.,
$p$--Adic quantum mechanics, Commun. Math. Phys.,
1989, Vol.123, pp.659--676
\bibitem{ADFV} \ Aref'eva I.Ya., Dragovic B., Frampton P., Volovich I.V.,
Wave function of the universe and $p$--adic gravity,
Mod. Phys. Lett. A, 1991, Vol.6, pp.4341--4358
\bibitem{cond-mat} \ Avetisov V.A., Bikulov A.H., Kozyrev S.V.,
Application of $p$--adic analysis to models of spontaneous
breaking of replica symmetry, Journal of Physics A, 1999, Vol.32,
pp.8785--8791,
http://xxx.lanl.gov/abs/cond-mat/9904360
\bibitem{PaSu} \ Parisi G., Sourlas N., $p$--Adic numbers and replica symmetry
breaking, http://xxx.lanl.gov/abs/cond-mat/9906095
\bibitem{Carlucci}\  Carlucci D.M., De Dominicis C.,
On the replica Fourier transform,
http://xxx.lanl.gov/abs/cond-mat/9709200
\bibitem{Carlucci1}\  De Dominicis C., Carlucci D.M., Temesvari T.,
Replica Fourier transform on ultrametric trees and block diagonalizing of
multireplica matrices,
J. Phys. I France, 1997, Vol.7, pp.105-115,
http://xxx.lanl.gov/abs/cond-mat/9703132
\bibitem{wavelets}\ Daubechies I., Orthonormal bases  of compactly supported
wavelets, Comm. Pure Appl. Math. Vol.41 (1988) p.906;
The wavelet transform, time frequency localization and signal analysis,
IEEE Trans. Inform. Theory Vol.36 (1990) p.961;
Ten Lectures on Wavelets, CBMS Lecture Notes Series, Philadelphia: SIAM,
1991
\end{thebibliography}
\end{document}